\documentclass[amssymb,prd,aps,amsmath,twocolumn,superscriptaddress,nofootinbib]{revtex4}
\usepackage{pslatex}
\usepackage{graphicx}
\usepackage{psfrag}

\usepackage[usenames]{color}
\usepackage{color,graphicx,epsfig,amsmath,amssymb}

\newcommand{\beq}{\begin{equation}}
\newcommand{\eeq}{\end{equation}}
\newcommand{\ben}{\begin{eqnarray}}
\newcommand{\een}{\end{eqnarray}}
\newcommand{\bi}{\begin{itemize}}
\newcommand{\ei}{\end{itemize}}

\newcommand{\ghost}[1]{ }

\newcommand{\myslash}[1]{\slash\!\!\!\!{#1}}

\begin{document}
\title{Signature of Sub GeV Dark Matter particles at LHC and TEVATRON}

%\title{LHC and TEVATRON experiments have the potential to constrain the MeV dark matter scenario}

\author{Daniel Albornoz V\'asquez}
\affiliation{LAPTH, Universit\'e de Savoie, CNRS, BP110, F-74941 Annecy-le-Vieux Cedex, France}
\email{albornoz@lapp.in2p3.fr}

\author{C\'eline B\oe hm}
\affiliation{LAPTH, Universit\'e de Savoie, CNRS, BP110, F-74941 Annecy-le-Vieux Cedex, France}
\email{celine.boehm@cern.ch}

\author{John Id\'arraga}
\affiliation{239 PSE Building
Department of Physics and Astronomy
York University
4700 Keele St
Toronto, ON M3J 1P3
Canada}
\email{idarraga@cern.ch}

\date{today}

\begin{abstract}
In this letter, we investigate the production of light dark matter particles at LHC in light of the model ($N=2$ SUSY inspired) proposed in Ref.~\cite{bf} and demonstrate that they will be copiously produced  if the colored messengers $F_q$ are lighter than 1 TeV. We expect up to $10^6$ events if $m_{F_q} \simeq 500$ GeV, assuming a $\sim$1 $fb^{-1}$ luminosity. In addition, we show that, even if $m_{F_q} > {\cal{O}}(1)$ TeV, searches for $F_q$ production at LHC are promising because a kinematical signature can be used to separate the signal from background. This signature is similar to that expected in supersymmetric scenarios. Hence, our study shows that most of the $m_{F_q}$ range could be constrained using LHC data. This should encourage further studies since they could infirm/confirm the MeV DM scenario.
\end{abstract}

\preprint{LAPTH-1370}

\maketitle

\section{Introduction}
\label{sec:intro}

%historic

It is often argued that an important motivation for carrying new physics searches at LHC and TEVATRON is the possibility of discovering 
 new particles that could enlight the nature of dark matter (DM). If DM is made of thermal particles, its relatively small abundance today 
 then guarantees a coupling to Standard Model particles and therefore suggests that it could be produced in high energy experiments. 

Unfortunately DM indirect detection at LHC and TEVATRON is not easy. The DM signatures that have been studied for a broad range of models (see for example  Supersymmetry, lepto-quarks, Kaluza-Klein, little Higgs) involve e.g. the production of leptons, di-leptons, jets as well as  missing energy \cite{Aad:2009wy}. 
In these frameworks, the DM mass is generally assumed to lie from a few GeV to a few TeV. Yet this is only a subset of all the DM possibilities that have been proposed in the literature. Besides, there is no evidence for such GeV-TeV DM particles in laboratory experiments so far (apart, perhaps, from \cite{Bernabei:2008yi,Ahmed:2009zw}), despite intensive searches.  In fact, direct detection experiments have now demonstrated that, if heavy neutral particles exist with a mass in the 100 GeV-TeV range, their interactions with ordinary matter must be small, e.g. \cite{Ahmed:2009zw,BarnabeHeider:2005pg,Lemrani:2006ec}, implying fine-tuned solutions (e.g. \cite{Ellis:1998kh}) to simultaneously explain the DM relic density.

In this context, it is worth investigating other DM candidates. Here we shall concentrate on 1-100 MeV  particles \cite{bens,bf} which were first introduced while studying the effect of DM interactions (whatever the nature of DM) on Large Scale Structure formation \cite{bfs,bs}. A subclass of these candidates (DM with a mass of a few MeV) has received a lot of attention after it was realized that it could explain the emission of 511 keV photons in the galactic centre while conventional astrophysical sources  generally fail to reproduce the observation \cite{511}.

The phenomenology and detectability of such MeV particles have been studied in detail \cite{bf,Boehm:2007na,Borodatchenkova:2005ct,Kahn:2007ru}. However, all previous studies 
assumed low energy experiments (where chances of detection seemed greater owing to the very small DM coupling to electrons \cite{ascasibar}). Here we show that similar studies for  high energy experiments are actually much more promising (on the reasonable assumption that DM is coupled to quarks).

The paper is organized as follows. In Sec.~\ref{sec:model}, we introduce our  model. In Sec.~\ref{sec:predictions}, we compute the number of events expected in an experiment with one $fb^{-1}$ luminosity. Motivated by previous analysis of similar final states, we study the kinematical characteristics associated to our signal and compare it with those of the background in  Sec.~\ref{sec:background}. We conclude in  Sec.~\ref{sec:conclusion}.

\section{The model \label{sec:model}}

Motivated by the model proposed in Ref.~\cite{bf} (based on $N=2$ supersymmetry), we  
consider a Lagrangian ${\cal{L}}$ containing a term 
\begin{equation}\
\label{lagrangian}
{\cal{L}} \supset S \ \delta_{ij} \ \bar{F}^{i} \ (c_l P_r + c_r P_l) \ f^j.
\end{equation}
where $\delta_{ij}$ denotes the Flavour indices,  $S$ an SU(2) singlet scalar field (our DM candidate),  $F$ a heavy fermion and $f$ a Standard Model (SM) particle.  

Such a term looks like the ``standard'' coupling between a sfermion, a SM fermion and a neutralino in $N=1$ supersymmetry, except that in our model $F$ is a spin 1/2 particle (associated with $f$) and $S$ is a neutral scalar. To preserve gauge invariance, $F_R$ 
is  a doublet of $SU(2)$ while $F_L$ is a singlet \cite{Fayet:1978ig,bf}. Hence terms like $S \bar{F_L} F_R$ or $H \bar{F_L} f_R$ (with $H$ the Higgs field) are absent from the Lagrangian since they are not gauge invariant.  Owing to $F$ quantum numbers, one could introduce other terms, such as $H \bar{F_L} F_R$ or $H \bar{F_L} f_l$. The former is expected to give a mass to the $F$ (hereafter denoted by $m_F$). However, the exact value of $m_F$ may arise from the existence of several Higgs doublet in the underlying theory and, potentially, contributions from soft symmetry breaking (like in $N=1$ SUSY).  The later term introduces a mixing mass matrix between the SM fermion and the $F$. However we assume the existence of a new symmetry ($M$-parity) \cite{Fayet:1978ig,bf} which kills such a term. 
 In principle, a phenomenological study of $N=2$ supersymmetry
 would require to write the full Lagrangian and would lead to all difficulties encountered already in $N=1$ SUSY. However we argue that, with this term only, one can already learn about such theories. As an illustration, we focus on the MeV DM scenario (which was based on this Lagrangian \cite{bf,511,ascasibar}). However a similar analysis could be done for heavier DM particles.
  
Let us now focus on the quark sector. For simplicity, we assume that all $F_q$ have a mass $m_{F_q} = m_F \neq m_q$ and all the $F_q$ couplings to their corresponding SM quark have the same values. This may cancel potential contributions to rare meson decays. In addition, we do not introduce any flavour mixing between the various $F$. CP phases are set to zero. Owing to these properties, there should not be any large FCNC contribution in this set up. We make similar assumptions for the leptonic sector, except that  $m_{F_{l}} \neq m_{F_{\nu}} (\neq m_{F_q})$. DM pair annihilation into neutrinos could thus insure the correct relic density \cite{Boehm:2008zz} (even though the relic density criteria may be alleviated by the assumption of two DM particles \cite{Boehm:2003ha}). 

We can now study the signatures associated with this scenario. To illustrate our purpose, we  consider couplings $c_{l,r}$ varying between $[0.3,3]$. Couplings above unity may, of course, appear rather unlikely as they may induce unseen anomalies in particle physics measurements (depending of $m_{F_q}$) and produce a very bright, yet extremely narrow,  monochromatic line through quark box diagrams at an energy $E= m_{dm}$ in our galaxy \cite{Boehm:2006gu}. However, for this analysis, 
we use these very large couplings as benchmark points to determine the typical values of the MeV DM coupling to quarks that can be probed at LHC. Note that the $F_q$ being ``colored'', they can be directly produced through gluon-gluon fusion.

\section{$F_q$ production cross sections in proton-proton collisions \label{sec:predictions}}

%We start by studying $F_q$ production at the parton level. 

\subsection{ $F_q$ production in $q q$, $q \bar{q}$, $g q$ collisions and  $gg$ fusion}

\begin{figure*}
	 \includegraphics[width=13cm]{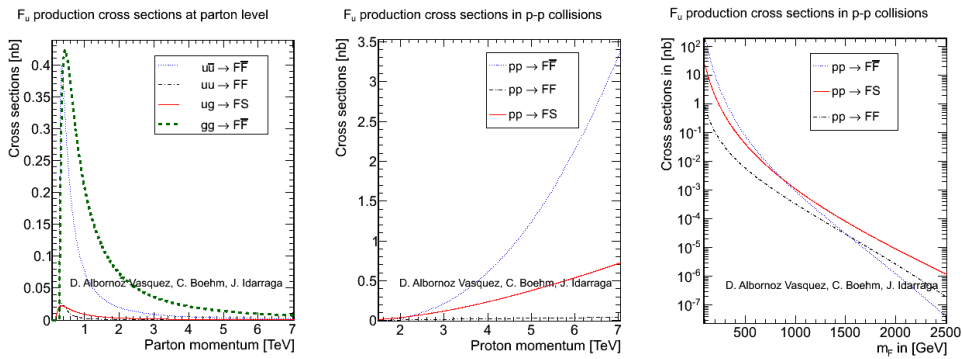}
	\caption{$F_q$ production cross sections. Left panel: at the parton level, assuming
          $m_F=300$ GeV, $c_{l,r} = 1$ and obtained by using the function $cs22$ implemented in
          micrOMEGAs. Middle panel:  at the proton level, with $m_F=300$ GeV, $c_{l,r} = 1$ and obtained by using hCollider. Right panel: dependence of the cross section with $m_{F_q}$. Same parameters as middle panel but $\sqrt{s} =10$ TeV.}
	\label{fig:parton}
\end{figure*}

The  $q \bar{q} \rightarrow F_q \bar{F_q}$ cross section involves a t and u-channel DM exchange as well as a s-channel gluon exchange. The $q q \rightarrow F_q F_q$ channel is similar but there is no s-channel gluon exchange \cite{Boehm:2009vn}. 
The $q g  \rightarrow F_q S$ process involves a t-channel $F_q$ exchange and a quark s-channel exchange while 
the $gg \rightarrow F_q \bar{F_q}$  is based on gluon fusion and t $+$ u-channel $F_q$ exchange. The behaviour of all these cross sections with respect to the parton energy is displayed in  Fig.~\ref{fig:parton} (left panel). 

Interestingly enough,  the $F_q$ production cross section through $qq$ collisions remain significant even for $m_{F_q} > 300$ GeV and $c_{l,r} < 3$. For example, it is about a few 30 pb for $m_{F_q} = 300$ GeV, $c_{l,r} =1$ and $\sqrt{s} \sim 2 \ m_{F_q}$. This can be understood by computing the matrix squared amplitude and writing the centre of mass energy as $s = 4 \ m_F^2 \ (1 + \epsilon^2)$ (where $\epsilon$ can be very large). For very large or very small values of $\epsilon$, one finds that $\sigma \propto 1/\epsilon$. But the cross section has  a maximum when $\epsilon$  satisfies:
\begin{eqnarray}
&&- 4 \ [ 7 (c_l^4+c_r^4) + 12 c_l^2 c_r^2 ] \epsilon^6 
- 16  \ [ 3 \ (c_l^4+c_r^4)   + 5 \ c_l^2 c_r^2  ] \epsilon^4 \nonumber\\
&&- 3  \ [ 5 \ (c_l^4+c_r^4)  + 8  c_l^2 c_r^2 ] \epsilon^2
+ 2 \ [ 3 (c_l^4+c_r^4) + 5 c_l^2 c_r^2 ]  = 0 \nonumber 
\end{eqnarray}
(assuming $\cos \theta =0$). Thus, if 
$c_l=c_r=1$, $\sigma_{qq \rightarrow F_q F_q}$ is maximal at 
$\epsilon \approx 0.47$ (or $\epsilon \approx 0.64$, if one integrates over $\cos \theta$ instead of setting it to zero).  We then find $\sigma \approx 30pb$, if $m_F = 300$ GeV. This cross section is therefore very large when the $F_q$ can be produced almost on-shell.

\subsection{$F_q$ production cross sections in p-p collisions}
 
To compute these cross sections, we used the function hCollider implemented in micrOMEGAs \cite{Belanger:2006is}. We have checked that our results were consistent with our analytical expressions (before convoluting with the parton distribution function) and the output from the VEGAS and Easy 2$\times$2 from CalcHEP \cite{Pukhov:2004ca}. Results from VEGAS and hCollider differ by a factor $\sim 1.5$ at high energy but this is due to the fact that we did not into take into account some QCD corrections in VEGAS.

Results are displayed in Fig.~\ref{fig:parton} (middle and right panel).
For not too large $c_{l,r}$ couplings, the dominant cross section corresponds to the $pp \rightarrow F_q \bar{F_q}$ process since it involves gluon fusion  and gluon exchange process. The latter can reach up to a few nb when $\sqrt{s} > 9$ TeV (assuming $m_{F_q} =300$ GeV) but it rapidly falls off with $m_{F_q}$. For $\sqrt{s} = 10$ TeV and $m_{F_q} =2$ TeV, it is only about 10 pb (see right panel of Fig.~\ref{fig:parton}).

\begin{figure*}
\includegraphics[width=4.5cm,angle=-90]{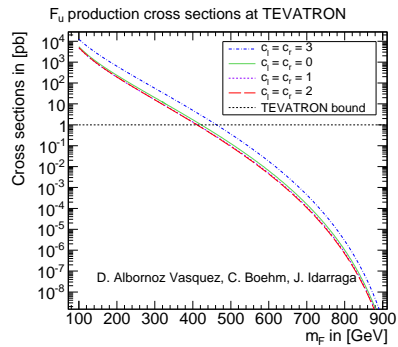}
\caption{$F_q$ production cross sections at Tevatron.}
\label{fig:tevatron}
\end{figure*}

These numbers suggest that TEVATRON may set a limit on $m_{F_q}$. 
Due to interference between the $S$ and gluon exchange diagrams, the $ p \bar{p} \rightarrow F_q \bar{F_q}$ production cross section is maximal for $c_{l}=c_{r} = 3$ and minimal for $c_{l}=c_{r} \simeq [1,2]$. For $c_{l}=c_{r} < 1$, 
the gluon exchange is the dominant process. Hence, the cross section is fixed by QCD couplings. 
As a result, we can set a limit on the only free parameter that is left, i.e. $m_{F_q}$. By analogy with LeptoQuark (jets$+$neutrinos) searches, we found that $ m_{F_q}$ should be greater than 450 GeV, see Fig.\ref{fig:tevatron}. However, this limit may also be 300 GeV, given that searches for this very model have not been implemented in TEVATRON experiments yet. Hence, to be ``conservative'', we shall use $m_{F_q} = 300$ GeV in the next section. Any greater value of $m_{F_q}$ will imply a smaller cross section.

%%%%%%%%%%%%%%%%%%%%%%%%%%%%%%%%%%%
%% Event signature and background
\section{Event signature and background \label{sec:background}}

\begin{figure*}
	\includegraphics[width=7cm]{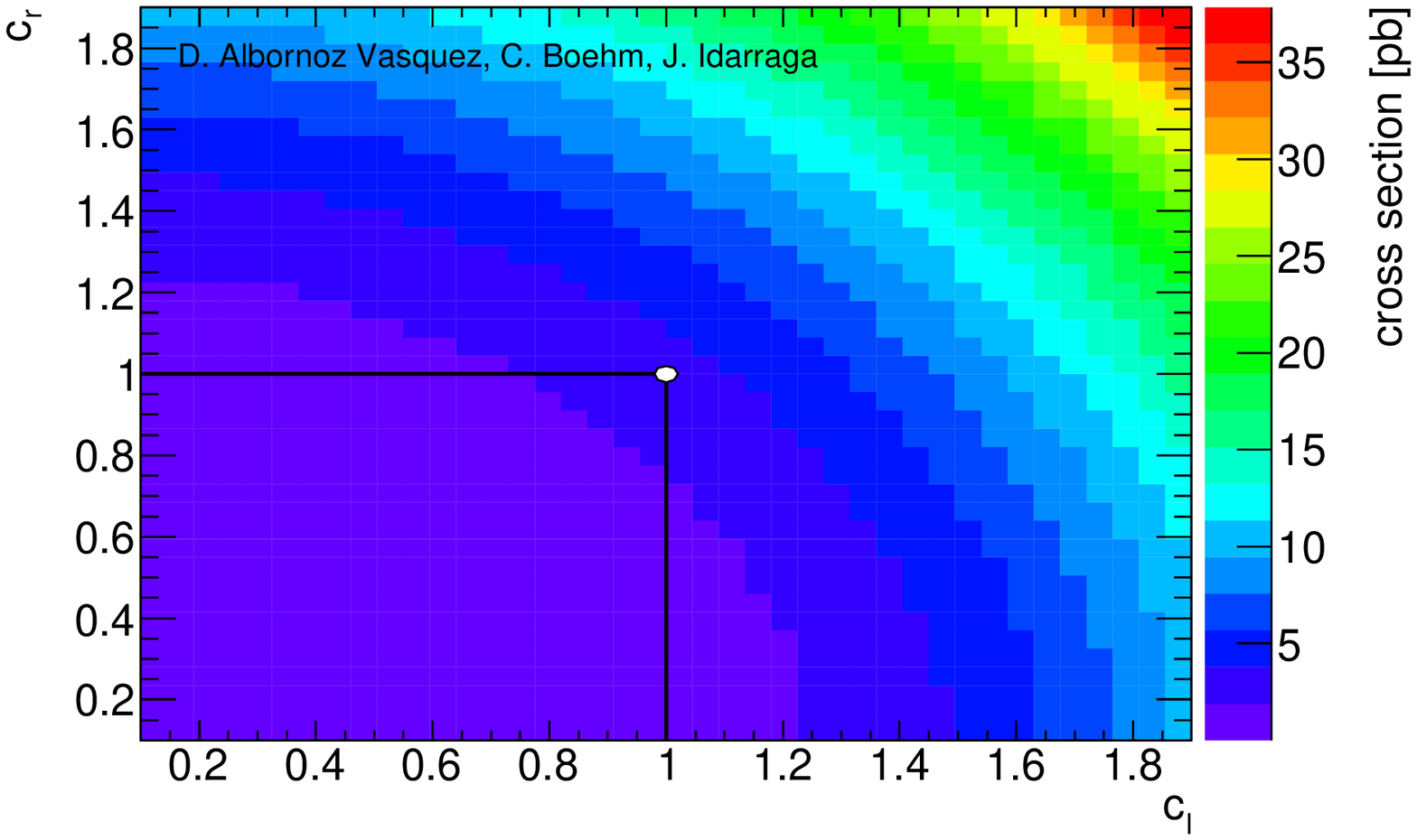}
	\includegraphics[width=6cm]{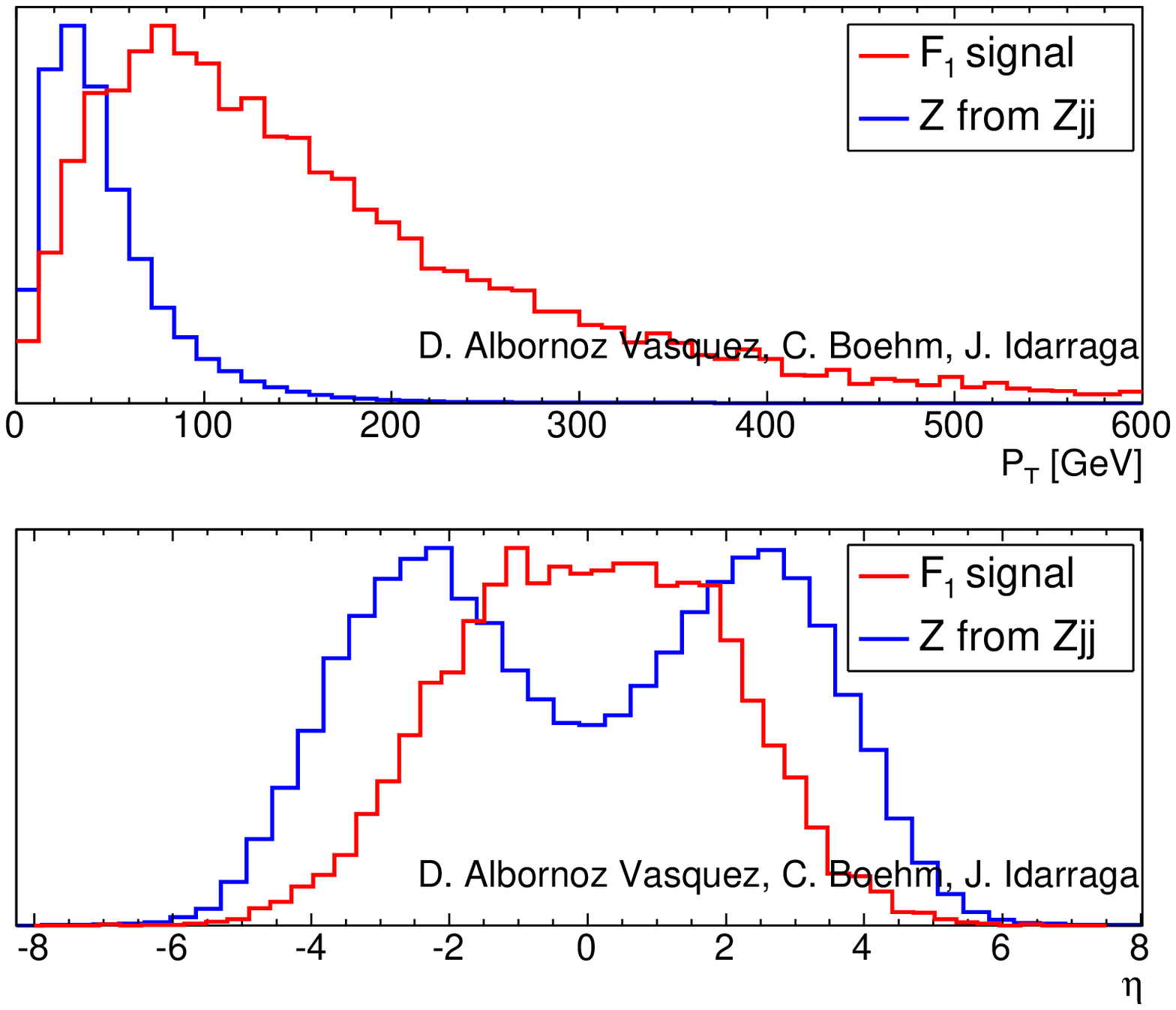}
	\caption{	$c_{l},c_{r}$ parameter's phase-space giving the
          production cross sections at the LHC at $\sqrt{s} = 7$ TeV
          (left).  Kinematic characteristics of the signal as opposed
          to those of the $Z$ boson in the associated $Z+$jets
          background (right).  In the figures (right), we have
          normalized the signal and background cross sections so that
          they become comparable.  We use $c_l=c_r = 1$, $m_F=300$
          GeV, $\sqrt{s} = 7$ TeV.}
	\label{fig:pp}
\end{figure*}

As shown in Sec.~\ref{sec:predictions}, $F_q$ should be produced
significantly in proton colliders at high energy.  In
Fig.~\ref{fig:pp} (left) we present the production cross sections as a
function of $c_{l},c_{r}$ parameters.  The production setup is $pp$
collisions at $\sqrt{s}=7$~TeV.  These are expected to be the initial
LHC conditions for the beginning of the physics program.  Our goal is
to study the kinematical properties of the signal and compare them to the
associated background in order to establish the discovery potential of
this signature at the LHC, particularly with the ATLAS
detector~\cite{Aad:2008zzm}.  ATLAS will be able to measure different
observables on the objects that compound the final state of our
signal, namely: jet identification and missing energy. Here a jet is
understood as the imprint left by the hadronization process of a high
energy quark in the detector material.

Since each $F_q$ decays into a jet $+$ missing energy
($\myslash{E}_{T}$), the associated background would be composed by:
First, $Z+$jets where $Z$ decays into two neutrinos.  Second,
$t\bar{t}$ where $t$ decays into $W,b$ and $W$ into $l,\nu$.  In cases
where the lepton falls out of the region of possible identification
($\eta>2.8$ for ATLAS), this signature can mimic the signal.
$t\bar{t}$ will be produced with an enormous cross section at the LHC
(see Table for all cross sections) and it has
to be taken into account in realistic simulations.  Even though,
previous studies have shown that it is not a real competitor for
signals with a final state composed by $2$ jets + $\myslash{E}_{T}$ as
shown in ~\cite{Aad:2009wy}(p.1595), or other combinations of
leptonic, hadronic plus $\myslash{E}_{T}$ final states as studied in
~\cite{jidarragaThesis}.  One of the advantages is that ATLAS has
$b$-tagging capability, and whenever an event clearly contains a
$b$-quark jet, we can reject it.  Third, the SM known processes
involving $WZ$, $ZZ$ and $WW$ production.  In the first and second
case it can perfectly mimic the signal when $W$ or $Z$ decays into two
jets and $Z$ decays into neutrinos.  It has been
proved~\cite{Aad:2009wy}(p.1595) that a cut based analysis based on
the kinematical properties of signal an backgrounds can eliminate this
background compared to signals down to a few fb.  The same techniques
are successful against the $Z+$~jets background mentioned above.
$WW$, like in the case of $t\bar{t}$, can only reproduce the signal
when one lepton falls out of the detectable region.  The cross
sections in the table below are production cross
sections, and do not include the branching ratio.  These cross
sections estimations have been produced with with the MadGraph
generator~\cite{Maltoni:2002qb} including pre-selection cuts
compatible with ATLAS calorimetry and tracking, according to those
used by the ATLAS collaboration in~\cite{Aad:2009wy}.

Even in pessimistic scenarios, where smaller values of the dark matter
couplings to quark (for example $c_l = c_r \simeq 0.3$) lead to cross
sections of the order of a few pb for $m_F < 200$ GeV (or down to 40 fb
for $m_F < 500$ GeV), the signal identification with this type of final
state is possible as shown in~\cite{Aad:2009wy}.  In Fig.\ref{fig:pp}
(right), we take the most competitive background, $Z$+jets (in this
case we consider only $Z+2$~jets at parton level produced with the
MadGraph generator~\cite{Maltoni:2002qb}) and plot the $p_T$ and
$\eta$ distributions against the same observables associated to one of
the $F_{q}$s in the signal.  The hardness of the $p_T$ distribution
for the signal, as opposed to that of the shown background, is a
typical characteristic of a two body decaying signatures at high
energy, and we include it here to show that we have successfully
implemented a MonteCarlo (MC) machine which allows us to study further
the kinematics of this signal.  We are going into a full simulation of
the associated final state using the Geant-4 simulation of the ATLAS
detector already thoroughly tested by the ATLAS
collaboration~\cite{Rimoldi:2001fm}~\cite{Aad:2009wy}.

Comparing the kinematical characteristics of our signal (which we have been 
able to study for the first time) with the signature of different signals which share the same final state (jets$+$missing energy) and are known as candidates for discovery at the LHC, we claim that current and forthcoming high energy experiments should have the ability to constrain the scenario presented in this letter.
\vspace{-0.7cm}
\begin{center}
$$
\begin{array}{|l|l|l|l|l|}\hline
\rm{Background}    &Z+n~jets  &t\bar{t} &WZ &ZZ \\ \hline
\rm{cross section} &     8~nb    &   10^3~pb     & 11~pb   &4~pb    \\
\rm{no} \ (\Gamma_{i}/\Gamma) \ \rm{fraction} & & & & \\ \hline
\end{array}
$$
\end{center}

%\begin{table*}
%\begin{tabular}{|l|c|}\hline
%Background     &   cross section \\ 
%               &   no $\Gamma_{i}/\Gamma$ fraction \\ \hline
%$Z+$n~jets     &       $8$~nb    \\ \hline
%$t\bar{t}$     &    $1000$~pb    \\ \hline
%WZ             &      $11$~pb    \\ \hline
%ZZ             &       $4$~pb    \\ \hline
%\end{tabular}
%\caption{Backgrounds cross section estimation.  In $Z+$n~jets, $1,2$
%  and $3$~jets samples are considered.}
%\label{table:backgrounds}
%\end{table*}

\section{Conclusion \label{sec:conclusion}}

In this letter, we investigate the case of DM particles coupled to SM quarks through a new type of colored charged fermions $F_q$ (as predicted in models \cite{bf} inspired from $N=2$ SUSY \cite{Fayet:1978ig}) and 
study $F_q$ production at LHC and TEVATRON. 
We show that, in the case of sub GeV DM, up to $10^6$ events could be produced in a collider 
with a 1 $fb^{-1}$ luminosity if $m_{F_q} \simeq 300$ GeV, and about $10^3$ events if $m_{F_q} \simeq 2$ TeV (with $\sqrt{s} = 10$ TeV). In addition we found that the kinematic characteristics of the signal in contrast to those of the associated backgrounds can be used on a based cut analysis on simulated or real data for mFq $>$ O(1). This indicates that searches for this type of couplings (at least in the case of light DM particles) should be possible at LHC and TEVATRON and should definitely motivate further studies in high energy experiments. 
One should also remember that a large production of unstable
colored $F_q$ particles could lead to a large production of
muons (after hadronization of the jets in the detector), which could be 
very useful to constrain such a scenario. This will be investigated in a forthcoming study. 
Finally, since the kinematical signature of this model is similar to that expected in some supersymmetric scenario, a study of the spin of the $F_q$ particle may be required to help for the identification of the dark matter.

\textbf{Acknowledgment:} We are grateful to G. Belanger, S. Davidson, P. Fayet, M. Kakizaki, A. Pukhov, R. Singh and F. Staub for 
illuminating discussions. 

\bibliography{LDMLHC3}

\end{document}